\documentstyle[12pt]{article}
\begin{document}
\newcommand{\mup}{{\mu}^{\prime}}
\newcommand{\nup}{{\nu}^{\prime}}
\newcommand{\muz}{{\mu}^{\prime \prime}}
\newcommand{\nuz}{{\nu}^{\prime \prime}}
\newcommand{\z}{\prime \prime}
\newcommand{\p}{\prime}
\newcommand{\rond}{\partial}

\begin{titlepage}
\vspace{-10mm}

\begin{center}
\begin{large}
{\bf CHAIN STRUCTURE \\ IN SYMPLECTIC ANALYSIS \\OF A CONSTRAINED SYSTEM   }\\
\end{large}
\vspace{5mm} {\bf M. Mojiri \footnote{e-mail:
mojiri@sepahan.iut.ac.ir}}, {\bf A. Shirzad \footnote{e-mail:
shirzad@ipm.ir}} \vspace{12pt}\\
{\it Department of  Physics, Isfahan University of Technology \\
Isfahan,  IRAN, \\
Institute for Studies in Theoretical Physics and Mathematics\\
P. O. Box: 5531, Tehran, 19395, IRAN.} \vspace{0.3cm}
\end{center}
\abstract{We show that the constraint structure in the chain by
chain method can be investigated within the symplectic analysis of
Faddeev-Jackiw formalism.
 \vfill}
\end{titlepage}

\section{Introduction}
In our previous paper \cite{shimoj} we showed that the traditional
constraint structure of Dirac formalism \cite{Dirac,BGP} can also
be obtained from symplectic analysis \cite{Sym1,Sym2,MONTA} of
Faddeev-Jackiw formalism \cite{FJ}. In the traditional Dirac
method, at each level of consistency, constraints divide to
first- and second-class. Hence, the consistency of second class
constraints at that level determines a number of  Lagrange
multipliers, while the consistency of first-class ones leads to
constraints of the next level. For this reason this method is
called {\it level by level}.

An alternative method, called {\it chain by chain}, has been
recently introduced \cite{lorshir}, in which constraints are
collected as first- and second- class chains. In this method the
consistency of each individual constraint in a definite chain
gives the next element of that chain, assuming that it is not the
terminating element. Such a chain structure possesses suitable
properties in constructing the gauge generating function
\cite{chaichian,shirshab} as well as the process of gauge fixing
\cite{fix}.

In this paper we show that the chain structure can also be derived
from the symplectic analysis. In other words, following the
singularity properties of symplectic two-form, one can find
suitable null-eigenvectors for it such that the resulting
constraints emerge in a chain structure. This would be done in
section (2). In section (3) we give more technical details about
terminating elements of the chains, together with discussing the
main properties of the chains (when they are first class and when
they are second class). Some examples are given in section (4)
and our concluding remarks are given in section (5).

\section{Chain Structure}
Consider a phase space with coordinates $y^i(i=1 , \dots ,2N)$
specified by the first order Lagrangian
\begin{equation}
L= a_i(y) \dot{y}^i-H(y) \label{a1}
\end{equation}
where $H(y)$ is the canonical Hamiltonian of the system. The
equations of motion read
\begin{equation}
f_{ij}\dot{y^j}=\partial_i H \label{a2}
\end{equation}
where
\begin{equation}
f_{ij} \equiv \partial_i a_j(y) - \partial_j a_i(y) \label{a3}
\end{equation}
It is called  {\it presymplectic tensor }. We denote it in matrix
notation as $f$. Suppose it is non-singular. Let $f^{ij}$ be the
components of the inverse, $f^{-1}$. Then from (\ref{a2}) we have
\begin{equation}
\dot{y}^i=\left\{ y^i,H \right\} \label{a3.5},
\end{equation}
where the Poisson bracket $\left\{\ ,\ {}\right\}$ is defined as
\begin{equation}
\left\{ F(y),G(y)\right\}=\partial_i F\partial_j G f^{ij}.
 \label{a4}
\end{equation}

Suppose we want to impose the set of primary constraints
$\Phi^{(1)}_{\mu}$ to the system. To do this, as stated in
\cite{shimoj}, one should add the consistency term
$\eta^{\mu}\dot{\Phi}^{0}_{\mu}$ to the canonical Hamiltonian and
extend the phase space to include the Lagrange  multipliers $\eta
^{\mu}$. This gives the next order Lagrangian
\begin{equation}
L^{(1)}=(a_i-\eta^{\mu}A_{\mu i})\dot{y}^i-H(y) \label{a5}
\end{equation}
where
\begin{equation}
A_{\mu i}=\rond _i \Phi^{(1)}_{\mu},
 \label{a6}
\end{equation}

 Considering $Y\equiv(y^i,\eta^{\mu})$ as coordinates, the
symplectic tensor $F$ reads
\begin{equation}
F = \left( \begin{array}{c|c}f& A \\
\hline -\tilde{A}  & 0
\end{array} \right).
 \label{a7}
\end{equation}
The equations of motion in the matrix notation are
\begin{equation}
F \dot{Y}= \partial H \label{a8}
\end{equation}

Using operations that keep the determinant invariant, it is easy
to show that
\begin{eqnarray}
\nonumber\det F&=&\det \left( \begin{array}{c|c} f& A \\
\hline 0 & \tilde{A}f^{-1}A
\end{array} \right) \nonumber \\ {} &=&(\det f)(\det \tilde{A} f^{-1}
A).
\label{a9}
\end{eqnarray}
Assuming $\det{f}\neq 0$, $F$ would be singular if $C \equiv
\tilde{A} f^{-1} A $ is singular. Using (\ref{a4}) and (\ref{a6})
we have
\begin{equation}
C_{\mu \nu}=\left\{\Phi^{(1)}_{\mu},\Phi^{(1)}_{\nu} \right\}.
\label{a10}
\end{equation}

Now we want to give a different approach compared to our previous
work in \cite{shimoj}. In that paper we investigated all null
eigenvectors of $F$. Here, considering the matrix $C_{\mu\nu}$ in
more detail, we concentrate only on the first row (and first
column). Suppose $\Phi^{(1)}_1$ has vanishing Poisson bracket with
all primary constraints, i.e. $\left\{
\Phi^{(1)}_1,\Phi^{(1)}_{\mu} \right\}\approx 0$. Then it is clear
that $F$ has the null eigenvector
\begin{equation}
\left(\partial _i \Phi^{(1)}_1 f^{ij},1,0,\cdots,0 \right).
 \label{a11}
\end{equation}
Multiplying both sides of (\ref{a8}) with (\ref{a11}) gives the
secondary constraint
\begin{equation}
\Phi^{(2)} _1 = \left\{ \Phi^{(1)}_{1},H \right\}. \label{a12}
\end{equation}

Next, we consider the consistency  of $\Phi^{(1)} _1 $ and add the
term $\eta^1_{(2)}\dot{\Phi}^{(2)}_1$ to the Hamiltonian. We array
the phase space coordinates as $$Y \equiv (y^i;\
\eta^1_{(1)},\eta^1_{(2)};\ \eta^2_{(1)},\cdots ,\eta^m_{(1)}).$$
Then the matrix $A$ at this stage reads
\begin{equation}
A \equiv \left(\rond \Phi^{(1)}_{1}, \rond \Phi^{(2)}_{1}\ ;\
\rond \Phi^{(1)}_{2},\cdots,\rond \Phi^{(1)}_{m}\right).
\label{a13}
\end{equation}
The matrix $C$ should also be improved to
\begin{equation}
C= \left( \begin{array}{c|c} C_{11}& C_{1\nu} \\
\hline C_{\mu 1} & C_{\mu \nu}
\end{array} \right)
\label{a14}
\end{equation}
where
\begin{equation}
C^{(nm)}_{\mu\nu} = \left\{ \Phi^{(n)}_{\mu} , \Phi^{(m)}_{\nu}
\right\} \label{a15}
\end{equation}
(So far , we have $m=1,2$ just for $\mu=1$. However, for $\mu>1$
only $m=1$ is present.)

Suppose again that $\Phi^{(2)}_{1}$ has vanishing Poisson brackets
with all primary constraints. Then a new null eigenvector would
emerge for $F$ as
\begin{equation}
\left(\rond_i \Phi^{(2)}_1f^{ij};\ 0,1;\ 0,\cdots,0 \right).
 \label{a16}
\end{equation}
Multiplying both sides of equations of motion (\ref{a8}) with
(\ref{a16}) gives the third level constraint
\begin{equation}
\Phi^{(3)} _1 = \left\{ \Phi^{(2)}_{1},H \right\}.
 \label{a17}
\end{equation}
The process will be continued by adding the term
$\eta^1_{(3)}\dot{\Phi}^{(3)}_1$ to the Hamiltonian and extend
the set of coordinates to include $\eta^{1}_{(3)}$ as well.
Proceeding in this way one produces the {\it constraint chain }
that begins with primary constraint $\Phi^{(1)}_1$(while keeping
the primary constraints $\Phi^{(1)}_2,\Phi^{(1)}_3, \cdots
,\Phi^{(1)}_m$ as they are). The first chain terminates if no new
constraint emerge at the terminating point, or if the singularity
of $F$ due to the constraints of first chain disappears. We
postpone the discussion on "how the chain would terminate" to the
next section.

The first chain terminated, one should proceed to the next chain
beginning with $\Phi^{(1)}_2$. At this stage the (rectangular)
matrix $A$ in (\ref{a7}) is as follows
\begin{equation}
A \equiv \left(\rond \Phi^{(1)}_{1},\cdots, \rond
\Phi^{(N_1)}_{1}\ ;\ \rond \Phi^{(1)}_{2},\cdots,\rond
\Phi^{(1)}_{m}\right). \label{a18}
\end{equation}
Suppose $\Phi^{(1)}_2$ has vanishing Poisson brackets with the
constraints of  the first chain and with $\Phi^{(1)}_{\mu}$ ,$\mu
> 2$ as well. It is easy to find that the symplectic matrix $F$
has the following null eigenvector
\begin{equation}
\left(\rond_i \Phi^{(1)}_2f^{ij}; \ \overbrace{0,\cdots,0}^{N_1}\
;\ 1,0,\cdots,0 \right).
 \label{a19}
\end{equation}
Multiplying with equation of motion gives the next constraint
$\Phi^{(2)}_2$ of the second chain. Then the second chain
(beginning with $\Phi^{(1)}_2$) can be knitted in the same way as
the first one. The second chain terminated, one can produce the
constraints of the third chain, and so on. In this way one can
construct the whole system of constraints (within the symplectic
analysis) as a collection of constraint chains.
\section{Terminal elements}
In this section we want to see how the constraint chains may
terminate. We will also discuss that whether the chains are first
or second class. For this reason we begin with a one-chain
system. In this case the "chain by chain method" coincides
exactly with level by level method investigated in the framework
of symplectic analysis in \cite{shimoj}. Suppose the chain,
beginning with $\Phi^{(1)}$, terminates after $N_1$ steps. Then
the rectangular matrix $A$ in (\ref{a7}) is
\begin{equation}
A\equiv\left( \rond \Phi^{(1)},\cdots,\rond \Phi^{(N_1)} \right)
 \label{a21}
\end{equation}
and the matrix elements of $C\equiv\tilde{A}f^{-1}A$ would be
\begin{equation}
\begin{array}{lr}
C^{(nm)}=\left\{ \Phi^{(n)},\Phi^{(m)} \right\}& n,m=1,\cdots,
N_1.
 \label{a22}
 \end{array}
\end{equation}
Using the Jacobbi identity it is possible to show that
\begin{equation}
C = \left( \begin{array}{ccccc} 0 & 0 & \cdots & 0 & C^{1N_1}
\\ 0 & 0 & \cdots & C^{2 (N_1-1)}&C^{2N_1} \\ \vdots & \vdots & { } & \vdots &\vdots \\
0 & C^{(N_1-1) 2} & \cdots & C^{(N_1-1) (N_1-1)} & C^{(N_1-1) N_1} \\
C^{N_11} & C^{N_12} & \cdots & C^{N_1 (N_1-1)} &
P^{N_1N_1}\end{array} \right) \label{a23}
\end{equation}
and
\begin{equation}
\det{C}=(-1)^{\frac{N_1}{2}} \det{\left(C^{1N_1} \right)^{N_1}}.
 \label{a24}
\end{equation}
If $C^{1N_1} \equiv\left\{ \Phi^{(1)}, \Phi^{(N_1)}\right\}\approx
0 $ then from (\ref{a9}) and (\ref{a24}) we have
$\det{F}\approx0$. In order that $\Phi^{(N_1)}$ be the terminal
element, multiplying the null eigenvector
\begin{equation}
\left(\rond_i \Phi^{(N_1)}f^{ij}; \
\overbrace{0,\cdots,1}^{N_1}\right)
 \label{a25}
\end{equation}
with the equations of motion (\ref{a8}) should give no new
constraint. This would be so if $\left\{\Phi^{N_1},H
\right\}\approx 0$.

If, on the other hand, $C^{1N_1} \neq 0$ we have $\det{C}\neq0$.
This means that all constraints of the chain are second class.
Moreover, $\det{F}\neq0$  and the symplectic two-form would be
invertible. As shown in \cite{MONTA} the inverse of $F$ can be
written as
\begin{equation}
F^{-1} = \left( \begin{array}{c|c}f^{-1}-f^{-1} A C^{-1}
\tilde{A} f^{-1} & -f^{-1} A C^{-1} \\ \hline C^{-1} \tilde{A}
f^{-1} & C^{-1} \end{array} \right). \label{a26}
\end{equation}
where $A$ is defined in (\ref{a21}). Then $F^{-1}$ defines a new
bracket between functions of the original phase space that is the
same as Dirac bracket\cite{shimoj}. Using (\ref{a26}), the
equations of motion (\ref{a8}) can be solved for
$\dot{\eta}_{(n)}$'s. Imposing the results to the Lagrangian and
adding a total derivative is equivalent to redefining the original
Hamiltonian as
\begin{equation}
H\longrightarrow H- \left\{ H, \Phi^{(n)}\right\}C_{nm} \Phi^{(m)}
\label{a27}
\end{equation}
where $C_{nm}$'s are elements of $C^{-1}$.

The whole procedure is in full agreement with Dirac approach in
the framework of chain by chain method\cite{lorshir}. That is, a
one chain system is completely first or second class. In the
latter case the system is called  {\it self-conjugate}, since the
matrix $C$ of the Poisson brackets of constraints of the chain,
i.e. (\ref{a23}), is non-singular.

Now let consider a system with two chains. This is  the case when
we are given two primary constraints, say $\Phi^{(1)}$ and
$\Psi^{(1)}$.

Considering the consistency of constraints in the framework of
chain by chain method, the authors of \cite{lorshir} have shown
that a double chain system may be of four categories. both first
class, one first class and one self conjugate second class, both
self conjugate second class, and finally two {\it cross conjugate}
second class. In the following we show that the same things may
emerge as the results of the symplectic analysis, provided that
one follows similar steps of chain by chain method. \vspace{5mm}

 \textbf{i) two first class chains} \vspace{5mm}

In the chain by chain method, this happens when the terminal
element of first chain, say $\Phi^{(N_1)} $, has vanishing Poisson
brackets with primary constraints $\Phi^{(1)}$ and $ \Psi^{(1)}$
as well as with Hamiltonian and then knitting the second chain,
the same thing is true for the terminal element $\Psi^{(N_2)}$.

In symplectic analysis, the procedure of knitting the first chain
is described more or less in section (2), leading to matrix $A$,
given in (\ref{a18}). One should notice that the singularity of
$F$ given in (\ref{a7}) is not removed at this step. In fact,
there are $N_1$ null eigenvectors corresponding to $N_1$ first
class constraints of the first chain. However, to begin knitting
the second chain, we keep these null eigenvectors as they are and
just search for {\it new null eigenvectors} corresponding to
second chain. In other words, to find the $(n+1) $th element of
the second chain we use the following  null eigenvector:
\begin{equation}
\left(\rond_i \Psi^{(n)}f^{ij}; \ \overbrace{0,\cdots,0}^{N_1}\ ;\
 \overbrace{0,\cdots,0}^{n-1}\ ;\ 1 \right)
 \label{a28}
\end{equation}
Multiplying this with equations of motion (\ref{a8}), as before,
gives the constraint
\begin{equation}
\Psi^{(n+1)} = \left\{ \Psi^{(n)},H \right\}.
 \label{a29}
\end{equation}
Knitting the $\Psi$-chain in this manner, we reach finally the
terminal element $\Psi^{(N_2)}$ which commute with $H$. We
remember that the final symplectic two-form possesses $N_1+N_2$
null eigenvectors.\vspace{5mm}

\textbf{ii) One first and one second class chain}\vspace{5mm}

 By now, it may have been clear to the reader that in
order to knit the constraint chains similar steps should be
followed in symplectic analysis and Dirac formalism. In
symplectic analysis we search for appropriate null eigenvectors
while in Dirac formalism we investigate directly  the consistency
conditions.

For the case under consideration suppose the $\Phi$-chain is first
class and the $\Psi$-chain is second class. In this case the
matrix $F$ has the following from
\begin{equation}
F= \left( \begin{array}{c|c|c} f & A_1 & A_2 \\
\hline -\tilde{A}_1 & 0 & 0 \\ \hline -\tilde{A}_2 & 0 & 0
\end{array} \right)
\label{a30}
\end{equation}
where
\begin{equation}
\begin{array}{lr}
A^{(n)}_{1i}=\rond_i \Phi^{(n)} & n=1, \cdots , N_1\\
A^{(n^{\p})}_{2i}=\rond_i \Psi^{(n^{\p})} & n^{\p}=1, \cdots , N_2
\end{array}
\label{a31}
\end{equation}

Concerning the part $A_1$(and $-\tilde{A}_1$), the matrix $F$ has
$N_1$ null eigenvectors as $$ \left(\rond_i \Phi^{(n)}f^{ij}; \
\overbrace{0,\cdots,1,\cdots,0}^{N_1}\ ;\
 \overbrace{0,\cdots,0}^{N_2} \right).$$
However,  since the constraints in $\Psi $-chain are second class
the singularity in part $A_2$ has been removed. If one omits the
columns and and rows of part $A_1$, the remaining matrix
\begin{equation}
F_{\textrm{inv}}= \left( \begin{array}{c|c} f & A_2 \\
\hline -\tilde{A}_2 & 0
\end{array} \right)
\label{a32}
\end{equation}
would be invertible, where its inverse is something similar to
(\ref{a26}).

If  conversely $\Phi$- chain were second class and the $\Psi$-
chain were first class, then $A_1$-part would be invertible and
$A_2$-part would have $N_2$ null eigenvectors. \vspace{5mm}

\textbf{iii) two self conjugate second class chains} \vspace{5mm}

In Dirac formalism this is the case  when both chains are second
class and the terminating element of each chain has non vanishing
Poisson bracket with the top element of the same chain. As shown
in \cite{lorshir} it is possible to redefine the Hamiltonian and
constraints in such a way that constraints of one chain commute
with the constraints of the other chain.

Following the steps of chain by chain method in symplectic
analysis, one finally reaches to the symplectic two-form shown in
(\ref{a30}) and (\ref{a31}); but this time it is invertible.
Considering the algebra of Poisson brackets one can show that the
inverse can be written as:
\begin{equation}
F^{-1} = \left( \begin{array}{c|c|c}f^{-1}-\sum_{i=1}^{2} f^{-1}
A_i C^{-1}_i \tilde{A}_i f^{-1} & -f^{-1} A_1 C^{-1}_1& -f^{-1}
A_2
C^{-1}_2 \\ \hline C^{-1}_1 \tilde{A}_1 f^{-1} & C^{-1}_1& 0\\
\hline C^{-1}_2 \tilde{A}_2 f^{-1} & 0 & C^{-1}_2
\end{array} \right). \label{a33}
\end{equation}
where
\begin{equation}
\begin{array}{lr}
C_1^{nm}= \left\{ \Phi^{(n)}, \Phi^{(m)}\right\}\\
C_2^{n^\p m^\p}= \left\{ \Psi^{(n^\p)}, \Psi^{(m^\p)}\right\}
 \label{a34}
\end{array}
\end{equation}
\vspace{5mm} \textbf{iv) two cross-conjugate second class
chains}\vspace{5mm}

In this case the chains have the same length and the terminal
element of each chain has non-vanishing Poisson bracket with the
top element of the other chain \cite{lorshir}. Each constraint in
$\Phi$-chain finds its conjugate in the $\Psi$-chain and
vice-versa.

Following all the steps needed to knit the chains, finally the
symplectic two-form $F$, is as written in (\ref{a30}) and
(\ref{a31}), noticing that $N_1=N_2$. Suppose $A \equiv
\left(A_1,A_2 \right)$ is a rectangular matrix with $2N_1$
columns. Then the inverse of $F$ would be as in (\ref{a26}). It
should be noted that the matrix $C$ of Poisson brackets in this
case is as follows
\begin{equation}
C= \left( \begin{array}{c|c} 0 & X \\
\hline -\tilde{X} & 0
\end{array} \right)
\label{a35}
\end{equation}
where
\begin{equation}
X^{nm}=\left\{ \Phi^{(n)}, \Psi^{(m)}\right\}.
 \label{a36}
\end{equation}

Analyzing  the general case, i.e. the multi-chain system, is more
or less similar to the two-chain system. The chains are collected
as first class, self-conjugate second class, and couples of
cross-conjugate second class chains. The essential points to
reach such a system of constraints can be understood from the
discussions given above, however, the details are complicated and
does not lead to any new point.

\section{Example}
As an example, consider the Lagrangian
\begin{equation}
L=\frac{1}{2}\left( \dot{x}-ay \right)^2
-byz-\frac{1}{2}cy^2-\frac{1}{2}dz^2
 \label{a37}
\end{equation}
where $x,y$ and $z$ are variables and $a,b,c$ and $d$ are parameters. The
primary constraints are $\Phi^{(1)}_1 =P_y$ and $\Phi^{(2)}_1
=P_z$. The canonical Hamiltonian is
\begin{equation}
H_c=\frac{1}{2}P_x ^2+ aP_xy+byz+\frac{1}{2}cy^2+\frac{1}{2}dz^2.
 \label{a38}
\end{equation}
The secondary constraints are $\Phi^{(1)}_2 =aP_x+bz+cy$ and
$\Phi^{(2)}_2 =by+dz$.

Different types of a two-chain system can be obtained  by
suitable choices of parameters. For $a=d=1$ and $b=c=0 $ the
chains are
\begin{equation}
\begin{array}{ccc}
\Phi^{(1)}_1=P_y&{ } & \Phi^{(2)}_1=P_z \\ \Phi^{(1)}_2=P_x & { }&
\Phi^{(2)}_2=z \label{a39}
\end{array}
.
\end{equation}
The first chain is first-class and the next one is second-chain.
For $a=b=0 $ and $c=d=1 $ we have two self-conjugate chain as
follows
\begin{equation}
\begin{array}{ccc}
\Phi^{(1)}_1=P_y & { }&\Phi^{(2)}_1=P_z \\ \Phi^{(1)}_2=y & { }&
\Phi^{(2)}_2=z \label{a40}
\end{array}
.
\end{equation}
Finally for $a=c=d=0 $ and $ b=1$ there are two cross-conjugate
chains as
\begin{equation}
\begin{array}{cc}
\Phi^{(1)}_1=P_y & \Phi^{(2)}_1=P_z \\ \Phi^{(1)}_2=z &
\Phi^{(2)}_2=y \label{a41}
\end{array}
.
\end{equation}
Now let discuss the above system in symplectic analysis. The first
order Lagrangian is
\begin{equation}
L=P_x \dot{x}+P_y \dot{y}+P_z \dot{z}-\left(\frac{1}{2}P_x ^2+
aP_xy+byz+\frac{1}{2}cy^2+\frac{1}{2}dz^2 \right)
 \label{a42}
\end{equation}
with the primary constraints $P_y $ and $P_z$. Suppose
$\left(y^1,\ldots ,y^6 \right)$ stand for
$\left(x,y,z,P_x,P_y,P_z \right)$. Adding the consistency term $
\eta^1\dot{P}_y,\eta^2\dot{P}_z $ to the Lagrangian (see
Eq.\ref{a5} ) the symplectic two-form $F$ for coordinates
$$Y \equiv \left( y^1,\ldots,y^6,\eta ^1 , \eta ^2 \right)$$ is similar to Eq. \ref{a7}
in which $f$ is a $6 \times 6$ symplectic matrix and
\begin{equation}
A=\left(\begin{array}{cc}
0&0\\0&0\\0&0\\0&0\\1&0\\0&1
\end{array} \right).
\label{a43}
\end{equation}
It has two null-eigenvectors as follows
\begin{equation}
\begin{array}{c}
v_1=\left(0,1,0,0,0,0\, ;\,1,0\right)\ \\
v_2=\left(0,0,1,0,0,0\, ;\,0,1\right).
\end{array}
\label{a44}
\end{equation}
According to the procedure given in this paper, we should
consider them one by one. Multiplying $v_1 $form left by the
equations of motion (\ref{a8}) gives the second level constraint
of the first chain as
\begin{equation}
\Phi^{(2)}_1=aP_x+bz+cy.
\label{a45}
\end{equation}
Now let consider different choice of parameters:

\textbf{i)}  suppose $a=d=1$ and $b=c=0$. Then $ \Phi^{(2)}_1=P_x$
which commute with primary constraints and Hamiltonian. So the
singularity of symplectic two-form due to first chain remains in
the system, and the first chain terminates at this step. Then
multiplying the null eigenvector $ v_2$ (see Eq. \ref{a44}) by
equations of motion (\ref{a8}) gives the constraint
$\Phi^{(2)}_2=z$ which is conjugate to $ \Phi^{(1)}_2=P_z$. In
this way the system of constraints (\ref{a39}) are reproduced
(one first-class and one second-class).

\textbf{ii)} Suppose $a=b=0$ and $c=d=1$. Then $ \Phi^{(2)}_1=y$
is conjugate to $ \Phi^{(1)}_1=P_y$. Adding the consistency term
$\eta^1_{(2)} \dot{y}$ to the Lagrangian, the singularity of
symplectic two-form due to first chain disappears. So the first
chain is second class and terminates at $ \Phi^{(2)}_1$. Again
from null eigenvector $v_2$ the second level constraint $
\Phi^{(2)}_2=z$ emerges $v_2$ which  is conjugate to
$\Phi^{(1)}_2=P_z$. The singularity due to second chain also
disappears by adding $\eta^2_{(2)} \dot{z} $ to the Lagrangian.
As observed, the system of two self-conjugate chain given in
(\ref{a40}) is derived from the symplectic analysis.

\textbf{iii)} Suppose $a=c=d=0$ and $b=1 $. Then $\Phi^{(2)}_1=z$
is conjugate to $\Phi^{(1)}_2=P_z$. According to the algorithm of
chain by chain method \cite{lorshir} in such a situation (when
the last element of a chain a chain does not commute with some
other primary constraint) one should begin to knit the next chain
and then investigate the consistency condition of both chains
simultaneously. In symplectic analysis, this should be done by
considering the next null eigenvector, i.e. $v_2$(see Eq.
\ref{a44}). Multiplying the equations of motion (\ref{a8}) by
$v_2$ gives the constraint $\Phi^{(2)}_2=y$. In this way the two
cross-conjugate chains (\ref{a41}) would be reproduced.

Adding the consistency term $$ \eta^1_{(1)}\dot{P}_y
+\eta^2_{(1)}\dot{P}_z +\eta^1_{(2)}\dot{z}
 +\eta^2_{(2)}\dot{y}$$ to the Lagrangian the $6 \times 4$ matrix
 $A$ in the symplectic two-form (\ref{a7}) takes the form
\begin{equation}
A=\left(\begin{array}{cccc}0&0&0&0\\0&0&0&1\\0&1&0&0\\0&0&0&0\\1&0&0&0\\0&0&1&0
\end{array} \right).
\label{a46}
\end{equation}
It is obvious that singularity of symplectic two-form is removed
and using $F^{-1}$ the brackets induced in phase space (see eq.
\ref{a4}) is the same as Dirac brackets due to second class
constraints $P_y,P_z,z,y$.

\section{Conclusion}
In this work we showed that the symplectic analysis is able to
construct the constraints in a chain structure. In fact, at each
stage one can play suitably with null-eigenvectors of the
symplectic tensor to produce any set of desired constraints. To
construct the constraint chains one should act with
null-eigenvectors one by one, such that at each stage only one
chain gains a new constraint.

We think that this work shows once again the essential equivalence
between symplectic analysis and the Dirac method. In fact, there
are some hidden calculation in Faddeev- Jackiw formalism and
symplectic analysis which is more or less equivalent to what done
in traditional Dirac method. We tried to show some of  theses
detailed calculations.

\end{document}